\documentclass[a4paper,12pt]{article}

\usepackage[utf8]{inputenc}
\usepackage[T1]{fontenc}
\usepackage{times}
\usepackage[colorlinks=true,linkcolor=black,citecolor=black,urlcolor=blue]{hyperref}
\usepackage{geometry}
\usepackage{titlesec}
\titleformat{\section}
{\bfseries\uppercase}{\thesection.}{1em}{}
\titleformat{\subsection}
{\bfseries}{\thesection.\thesubsection.}{1em}{}

\usepackage{graphicx} 
\usepackage{multirow} 
\usepackage{cite}
\usepackage{breakurl}
\usepackage{indentfirst}
\usepackage{amsmath, amssymb, amsfonts, bm}
\usepackage{txfonts}
\usepackage{enumitem}
\usepackage{xcolor}
\usepackage{enumitem}
\usepackage{wallpaper}

\titlespacing*{\subsection}{0pt}{1.5em}{0.2em}

\renewcommand\eqref[1]{Equation~\ref{#1}}

\renewcommand{\thesection}{\arabic{section}}
\renewcommand{\thesubsection}{\arabic{subsection}}

\newlength{\bibitemsep}
\newlength{\bibparskip}
\let\oldthebibliography\thebibliography
\renewcommand\thebibliography[1]{%
  \oldthebibliography{#1}%
  \setlength{\parskip}{\bibitemsep}%
  \setlength{\itemsep}{\bibparskip}%
}

\usepackage{fancyhdr}
\usepackage{xcolor} 
\usepackage{lastpage} 

\definecolor{mygrey}{gray}{0.5}

\usepackage{fancyhdr}
\fancypagestyle{plain}{%
  \fancyhf{} 
}

\begin{document}
\thispagestyle{plain}

\begin{center}
	\includegraphics[width=\linewidth]{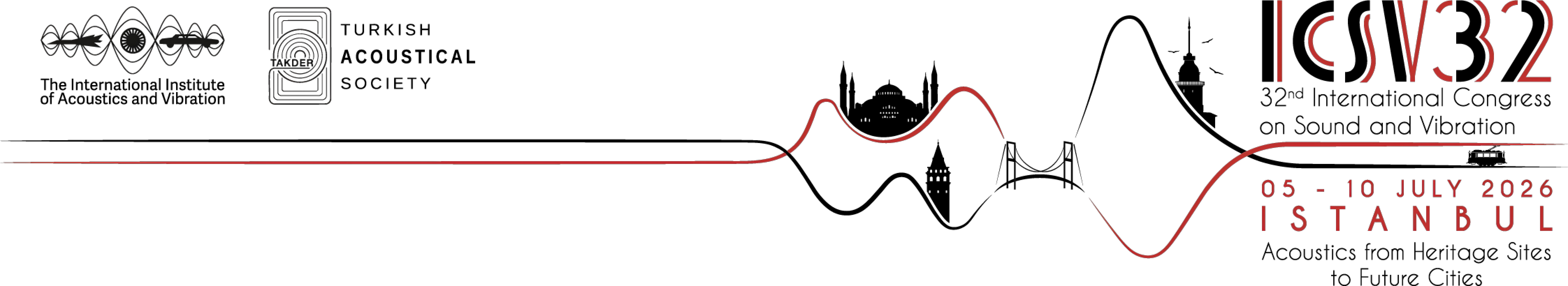}
\end{center}
\vskip.5cm

\begin{flushleft}
\fontsize{16}{20}\selectfont\bfseries
\color{black} { \textbf{Model-Agnostic Meta-Learning Initialization for Distributed Multichannel Active Noise Control} }%
\end{flushleft}
\vskip1cm

\renewcommand\baselinestretch{1}
\begin{flushleft}

 \textbf{Xiaoyi Shen$^{1}$} \\[2pt]
  {\footnotesize $^{1}$\textit{State Key Laboratory of Acoustics and Marine Information, Institute of Acoustics, Chinese Academy of Sciences, Beijing, 100190, China}} \\
  { \textit{email: \href{mailto:shenxiaoyi@mail.ioa.ac.cn}{shenxiaoyi@mail.ioa.ac.cn}}}\\[10pt]
  \textbf{Junwei Ji$^{2}$, Woon-Seng Gan$^{2}$} \\[2pt]
  { \footnotesize $^{2}$\textit{School of Electrical and Electronic Engineer, Nanyang Technological University, 639798, Singapore}}\\[10pt]
  \textbf{Dongyuan Shi$^{3}$} \\[2pt]
  {\footnotesize  $^{3}$\textit{Center of Intelligent Acoustics and Immersive Communications, Northwestern Polytechnical University, Xi'an, 710072, China}}\\[10pt]
  \textbf{Jun Yang$^{1,4}$} \\[2pt]
 {\footnotesize  $^{4}$\textit{School of Electronic, Electrical and Communication Engineering, University of Chinese Academy of Sciences, Beijing, 100049, China}} 
 %


\end{flushleft}

\textbf{\centerline{ABSTRACT}}\\
\textit{Distributed multichannel active noise control (DMCANC) has emerged as a scalable framework for large-area noise reduction, where multiple nodes operate local single-channel ANC controllers and exchange essential information to achieve global control. A key limitation of existing DMCANC implementations lies in their reliance on zero or random initialization, which leads to slow convergence of adaptive filters and restricts the efficiency of inter-node collaboration. To address this issue, this paper introduces a model-agnostic meta-learning (MAML)–based initialization strategy for DMCANC. By aggregating heterogeneous acoustic characteristics across nodes—including primary and secondary paths—a MAML framework is trained to learn an initialization that generalizes effectively across distributed ANC systems. The MAML initialization is then deployed to all nodes to improve convergence speed under both stationary and time-varying noise conditions. Numerical simulations applied on broadband and real-world noise demonstrate that the proposed algorithms achieves substantially faster convergence and improved noise reduction performance compared with conventional DMCANC, highlighting the potential of MAML initialization as an effective method for large-scale ANC.\\
}


\newpage
\section{INTRODUCTION}\label{Sec_1}
Active noise control (ANC) has been widely used as an effective technique for attenuating low-frequency noise by generating anti-noise signals with the same amplitude and opposite phase ~\cite{Kuo1999ANC,hansen1999understanding,Elliott2001SPAC}. Benefiting from adaptive filtering algorithms such as the filtered-x least mean square (FxLMS)~\cite{haykin2002adaptive,shi2018partial,lam2018active}, ANC systems are capable of operating in dynamic environments. With the increasing demand for large-area noise reduction, multichannel ANC (MCANC) systems have been extensively studied~\cite{chang2020active,SHEN2025112415}. Conventional MCANC systems are typically implemented in a centralized manner, where a single controller processes all sensor and actuator signals~\cite{shen2023momentum,gan2023practical}. Although such centralized schemes can achieve optimal performance, they suffer from high computational complexity and limited scalability. To address these issues, distributed multichannel ANC (DMCANC) systems have been proposed~\cite{Ji2023Distributed,Li2023Distributed}, in which multiple nodes independently perform local adaptation while exchanging information to achieve global noise reduction. This distributed architecture provides improved scalability, flexibility, and suitability for wireless implementation~\cite{SHEN2022117300,shen2024survey}.

Recent advances in DMCANC have focused on improving system robustness and practical feasibility under non-ideal communication conditions~\cite{Chu2020DiffusionANC,Li2023AugmentedDiffusion,zhou2025distributed}. The mixed-gradient FxLMS (MGDFxLMS) algorithm  applied compensation filters to combine the received gradients and mitigate crosstalk effects between each node~\cite{Ji2025MGDFXLMS}. MGDFxLMS has been proved haveing equivalent noise reduction performance to the centralized ANC system under ideal communication conditions. Furthermore, intermittent communication strategies have been introduced to DMCANC (IC-DMCANC) to reduce communication burden and computational complexity, while maintaining performance close to centralized systems~\cite{JI2026114024}. In such frameworks, each node typically runs a local adaptive algorithm, and global coordination is achieved through periodic information exchange. Despite these improvements, most existing DMCANC methods rely on zero or random initialization of control filters~\cite{shi2022selective,shi2023transferable}.

However, the initialization of adaptive filters plays a critical role in determining the convergence behavior of ANC systems~\cite{kajikawa2012recent,shi2024behind,luo2023deep}. In distributed settings, poor initialization not only slows down local adaptation but also limits the effectiveness of inter-node collaboration, especially under time-varying noise conditions~\cite{Ji2025SBWCFXLMS}. This issue becomes more pronounced in large-scale DMCANC systems, where heterogeneous acoustic paths and dynamic environments further complicate the convergence process~\cite{shen2022adaptive}.

To address this limitation, we propose a model-agnostic meta-learning (MAML)-based initialization framework for IC-DMCANC~\cite{finn2017model,zintgraf2019fast,shi2021fast,11021385}. Unlike conventional approaches that treat each ANC node independently, the proposed method leverages meta-learning to extract shared knowledge from heterogeneous acoustic environments, including variations in primary and secondary paths. By training across multiple tasks, the MAML framework learns a generalized initialization that enables rapid adaptation when deployed in unseen DMCANC scenarios. The proposed method is designed to be compatible with existing distributed ANC architectures. The learned initialization can be directly applied to each node as the initial control filter of the adaptive algorithms, thereby significantly accelerating convergence without altering the underlying control structure. Extensive simulations under broadband and real-recorded noise conditions demonstrate that the proposed approach achieves faster convergence and improved noise reduction performance compared with conventional DMCANC methods.

The remainder of this paper is organized as follows. Section~\ref{Sec_2} introduces the DMCANC system and reviews relevant adaptive algorithms and presents the proposed MAML-based initialization framework.  Section ~\ref{Sec_3}  provides performance analysis, followed by numerical simulations. Finally, conclusions are drawn in  Section ~\ref{Sec_4} .

\section{PROPOSED METHOD}\label{Sec_2}
\begin{figure}
    \centering
    \includegraphics[width=0.8\linewidth]{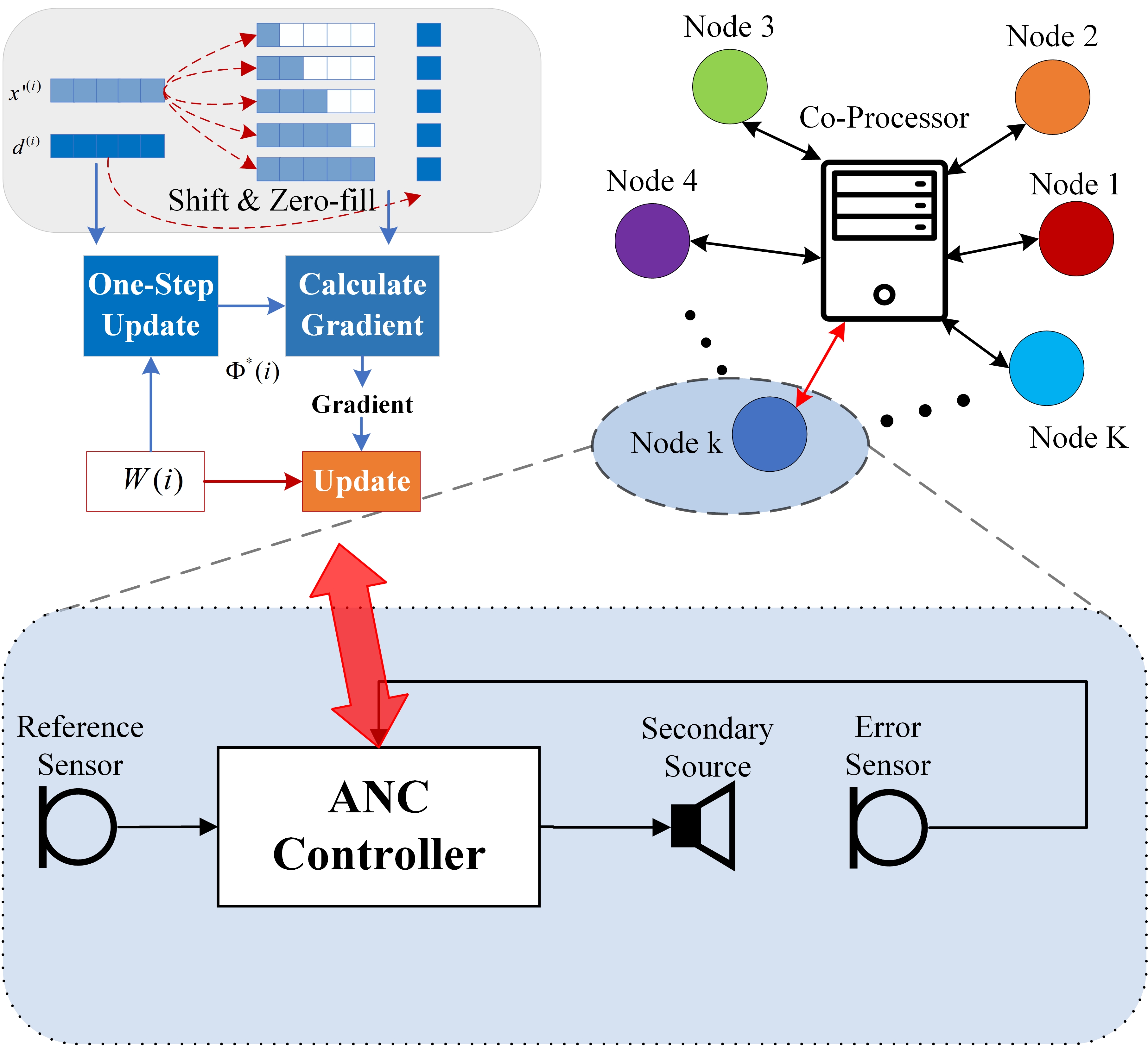}
    \caption{The block diagram of implemetation for IC-DMCANC. }
    \label{fig_icdmcanc}
\end{figure}

In order to imrove the convergence speed of IC-DMCANC, MAML is applied to pretrain the optimal initial control filter for each node of IC-DMCANC, which is shown in Fig.\ref{fig_icdmcanc}.

\subsection{ Distributed Multichannel ANC with Intermittent Communication }
DMCANC is composed of $K$ nodes, in the $k$-th node, the control signal can be calculated as:
\begin{equation}
     y_k(n) = \mathbf{w}_k^\mathrm{T}(n)\mathbf{x}(n), \quad k = 1,2,...,K,
\end{equation}
where $ y_k(n)$ denotes the control signal of the $k$-th node, and $\mathbf{x}(n)$ is the reference signal captured by the reference sensor.
$ \mathbf{w}_k(n) = [w_{k,1}(n) ,w_{k,2}(n) , \cdots , w_{k,L}(n)]^\mathrm{T}$ represents the weights of adaptive filter in the $k$-th node with the length of $L$, which is updated by steepest gradient descent method:
\begin{equation}\label{eq:WCFxLMS}
    \mathbf{w}_k(n+1) = \mathbf{w}_k(n) + \mu_k \mathbf{x}_{kk}'(n)e_k(n) + \mu_k \alpha\left[\widetilde{\mathbf{w}}_k-\mathbf{w}_k(n)\right].
\end{equation}
where $\mu_k$ and $\alpha$ denote the step size  and penalty factor for the weight update equation of $k$-th node.  $\widetilde{\mathbf{w}}_k$ stands for the center point of the range to which the control filter is restricted. $e_k(n)$ is the error signal of the $k$-th node and $\mathbf{x}_{kk}'(n)$ is the filtered reference signal calculated by
\begin{equation}
     \mathbf{x}_{kk}'(n) = \hat{s}_{kk}(n)*\mathbf{x}(n),
\end{equation}
where $\hat{s}_{kk}(n)$ denotes the estimated self-secondary path $s_{kk}(n)$ with length $L_s$. The compensation filters are introduced to make up for the difference between the node's self-secondary path and inter-node cross-secondary path. The compensation filters  $\mathbf{c}_{mk}(n)$  can be estimated offline prior to the operation of the ANC system: 
\begin{equation}\label{eq:compensationfilter}
    \mathbf{c}_{mk}(n+1) = \mathbf{c}_{mk}(n) + \mu_{c}\mathbf{{v}'}(n)e_m(n),
\end{equation}
where $\mathbf{{v}'}(n)$ denotes to the filtered WGN vector and $e_m(n)$ is given by
\begin{equation}\label{eq:compenerror}
    e_{m}(n) = v'_{k}(n)-v'_{m}(n),
\end{equation}
and $v'_k(n)$ and $v'_m(n)$ is generated by
\begin{equation}\label{eq:desiredandanti}
    \begin{cases}
        {v}'_k(n) = & v(n) * s_{mk}(n), \\
        {v}'_m(n) = & \mathbf{c}_{mk}^{\mathrm{T}}(n)\mathbf{v}(n) * s_{mm}(n),
    \end{cases}
\end{equation}
 in which $v(n)$ is white Gaussian noise and $s_{mk}(n)$ is cross-secondary path.

Intermittent communication (IC) strategy exchange data only at non-continuous, scheduled intervals rather than at every sampling point:
\begin{equation}\label{eq:WCFxLMS_N}
    \mathbf{w}_k(n+1) = \mathbf{w}_k(n-N+1) + \sum^{N-1}_{t=0}\mu_k\mathbf{x}'_{kk}(n-t)e_k(n-t)  + \sum^{N-1}_{t=0}\mu_k \alpha\left[\widetilde{\mathbf{w}}_k-\mathbf{w}_k(n-t)\right],
\end{equation}
which accumulates the algorithm’s gradient information over the $N$ steps. The new central control filter can be approximated as 
\begin{equation}\label{eq:MWD}
    \widetilde{\mathbf{w}}_{k}^{(r+1)} \approx   \widetilde{\mathbf{w}}_k^{(r)} + {\phi_k}(n) + \sum_{\substack{m=1\\ m\neq k}}^{K} \phi_m(n)*c_{mk}(n),
\end{equation}
where
\begin{equation}\label{eq:WD}
{\phi_k}(n) = \mathbf{w}_k(n+1) - \widetilde{\mathbf{w}}_k^{(r)}.
\end{equation}

\subsection{Model-Agnostic Meta-Learning Initialization}
In order to accelerate the noise reduction process, we applied Model-Agnostic Meta-Learning (MAML) initialization to obtain the optimal initial coefficients $\mathbf{w}^o_{k}$ for $k$-th ANC node.  The dataset is collected seperately using primary path and seconadry path for this node. Here we use one node as example. The dataset used for $k$-th node coefficient initialization consists of randomly sampled disturbance signals $\{\mathbf{d}^{(1)}_{k}(n),\mathbf{d}^{(2)}_{k}(n),\cdots\}$ and their corresponding filtered reference signal tracks $\{{\mathbf{x}_{kk}^\prime}^{(1)}(n),{\mathbf{x}_{kk}^\prime}^{(2)}(n),\cdots\}$ in different ANC configuration settings. The error signal of the $j$-th within-task training ${e_{k}^\star}^{(j)} (n)$ is obtained from 
\begin{equation}\label{eq_14}
	{e_{k}^\star}^{(j)} (n) = d^{(j)}_{k}(n)-[{\mathbf{x'}^{(j)}_{kk}}]^{\mathrm{T}}(n)\bm{\varrho}^{(j)}_k,
\end{equation} 
where $\bm{\varrho}^{(j)}_k$ represents the $j$-th initial value of the control filter. It is presumed that the best initial control filter requires only a
single update to reach the optimal solution $\mathbf{w}^o_{k}$:
\begin{equation} \label{feedforward}
    {\mathbf{w}^o_{k}}^{(j)} =\bm{\varrho}^{(j)}_k -\beta {e_{k}^\star}^{(j)} (n) {\mathbf{x}'_{kk}}^{(j)} (n),
\end{equation}
in which $\beta$ denotes the step size for the updation process in the within-task training. 

In $j$-th within-task testing, the testing data pair is given by 
\begin{equation}
\begin{small}
    \begin{cases}
    {d_{k}^\dagger}^{(j)}(m) &= {d_{k}}^{(j)}(n), \\
    {\mathbf{x}_{kk}^\dagger}^{(j)}(m-i) = &\begin{bmatrix}
		{x'_{kk}}^{(j)}(n-i) & {x'_{kk}}^{(j)}(n-i-1) &\cdots & \mathbf{0}_{1\times i}
	\end{bmatrix}^\mathrm{T},
    \end{cases}        
\end{small}
\end{equation}
where $i = 1,2,\cdots, L$, and the $j$-th within-task testing error ${e_{k}^\dagger}^{(j)}(m)$ is obtained from  
\begin{equation}\label{eq_6}
{e_{k}^\dagger}^{(j)}(m-i)={d_{k}^\dagger}^{(j)}(m-i)-[{\mathbf{x}_{kk}^\dagger}^{(j)}(m-i)]^\mathrm{T}{\mathbf{w}^o_k}^{(j)}.
\end{equation}
According to the gradient descent method,  the recursive formula $\bm{\varrho}_k$ is derived as 
\begin{equation}
\bm{\varrho}^{(j+1)}_k = \bm{\varrho}^{(j)}_k +\epsilon\sum_{i=0}^{L-1}\lambda^i {e^\dagger_{k}}^{(j)}(m-i){\mathbf{x}_{kk}^\dagger}^{(j)}(m-i),    
\end{equation}
where $\lambda \in \left[0,1 \right]$ stands for the forgetting factor and $\epsilon \in (0,1)$ denotes the learning rate. After within-task training and testing, the optimal initial value $\mathbf{w}^o_{k}$ obtained in \eqref{feedforward} is applued for $k$-th node as the initial coefficients to further update $\mathbf{w}_k(n)$ .

\section{SIMULATION RESULTS}\label{Sec_3}
\begin{figure}[h!]
    \centering
    \includegraphics[width=0.8\linewidth]{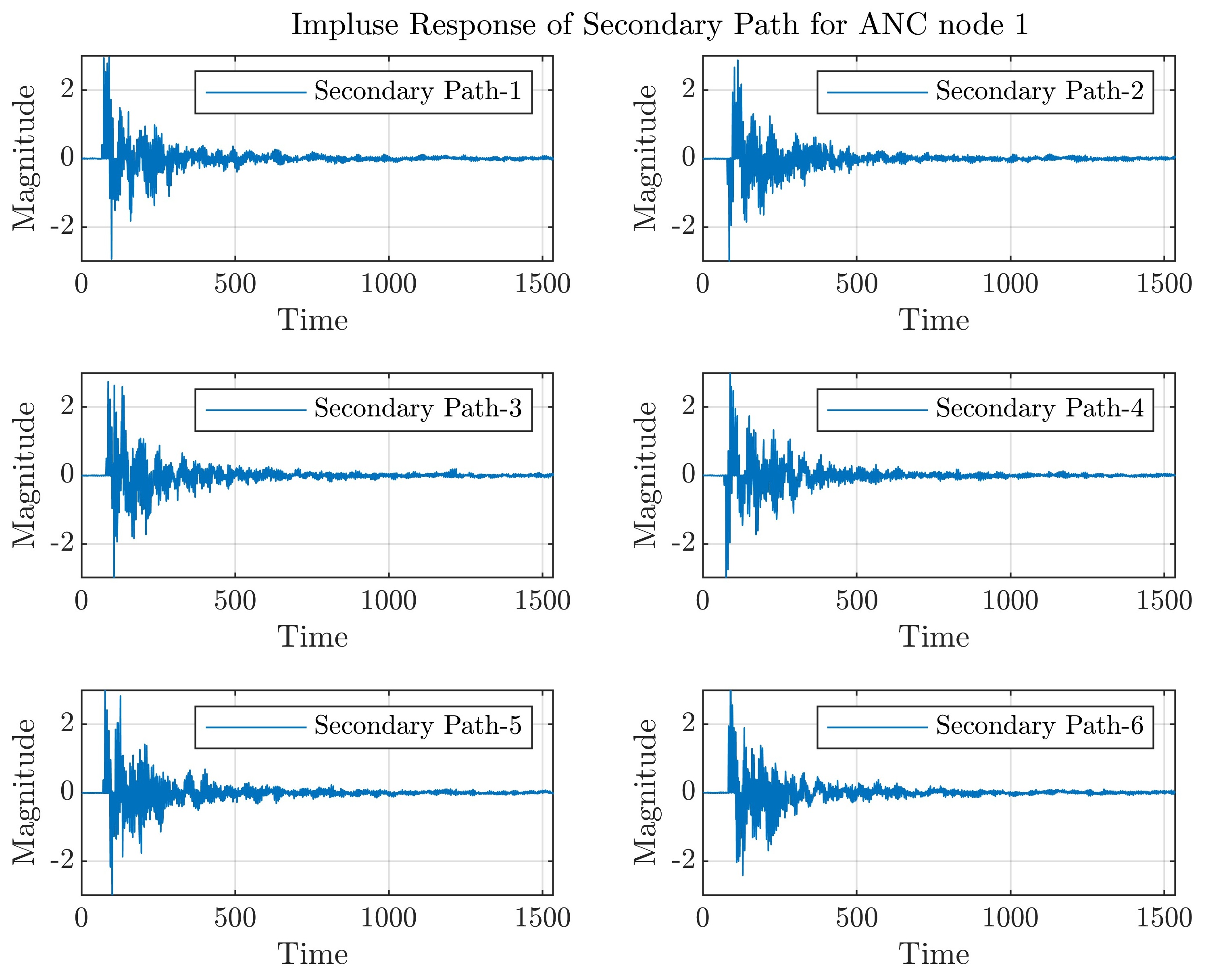}
    \caption{Impluse response of estimated secondary path for ANC node 1.}
    \label{fig_secondarypath}
\end{figure}

Simulations were conducted to validate the noise reduction performance of the proposed MAML-ICDMCANC framework, implemented with six nodes as shown in Fig.~1. The primary and secondary paths were measured in a real acoustic environment using an acoustic chamber. The impulse response of the secondary path for ANC node 1 is illustrated in Fig.~\ref{fig_secondarypath}. The lengths of the secondary path, compensation filter, and control filter were set to $256$, $33$, and $512$ taps, respectively. The sampling frequency for all algorithms was set at $16$ kHz. For comparison, centralized implementations and MGDFxLMS, as well as the IC-DMCANC algorithm, were also evaluated.

In the MAML initialization stage, broadband noise signals with frequency ranges of $100$–$1200$ Hz, $800$–$1500$ Hz, and $1200$–$2000$ Hz were used to train the optimal initial coefficients. The training data pairs were generated by passing the signals through the primary and secondary paths. The dataset was split into $70\%$ for training and $30\%$ for validation. The learned initialization was then applied to the adaptive filters under different noise conditions, while it was initialized as a zero vector for other compared algorithms.
\subsection{Tonal noise cancellation}
\begin{figure}[h!]
    \centering
    \includegraphics[width=0.8\linewidth,height = 9cm]{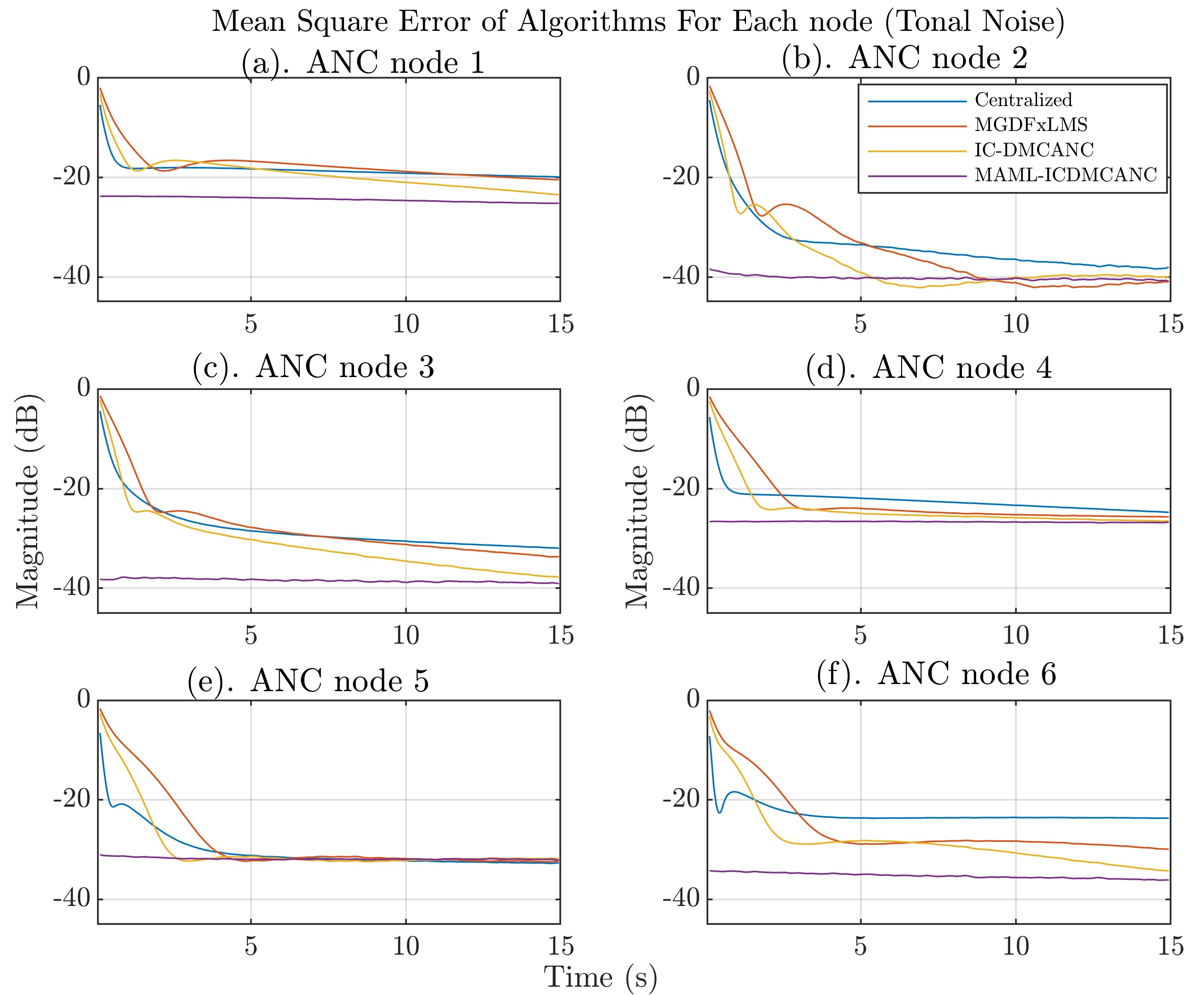}
    \caption{Mean square error of different algorithms for tonal noise with each ANC node: (a). Mean square error in ANC node 1. (b).  Mean square error in ANC node 2. (c).  Mean square error in ANC node 3. (d).  Mean square error in ANC node 4. (e).  Mean square error in ANC node 5. (f). Mean square error in ANC node 6.}
    \label{fig_mse1}
\end{figure}
\begin{figure}[h!]
    \centering
    \includegraphics[width=0.8\linewidth,height = 9cm]{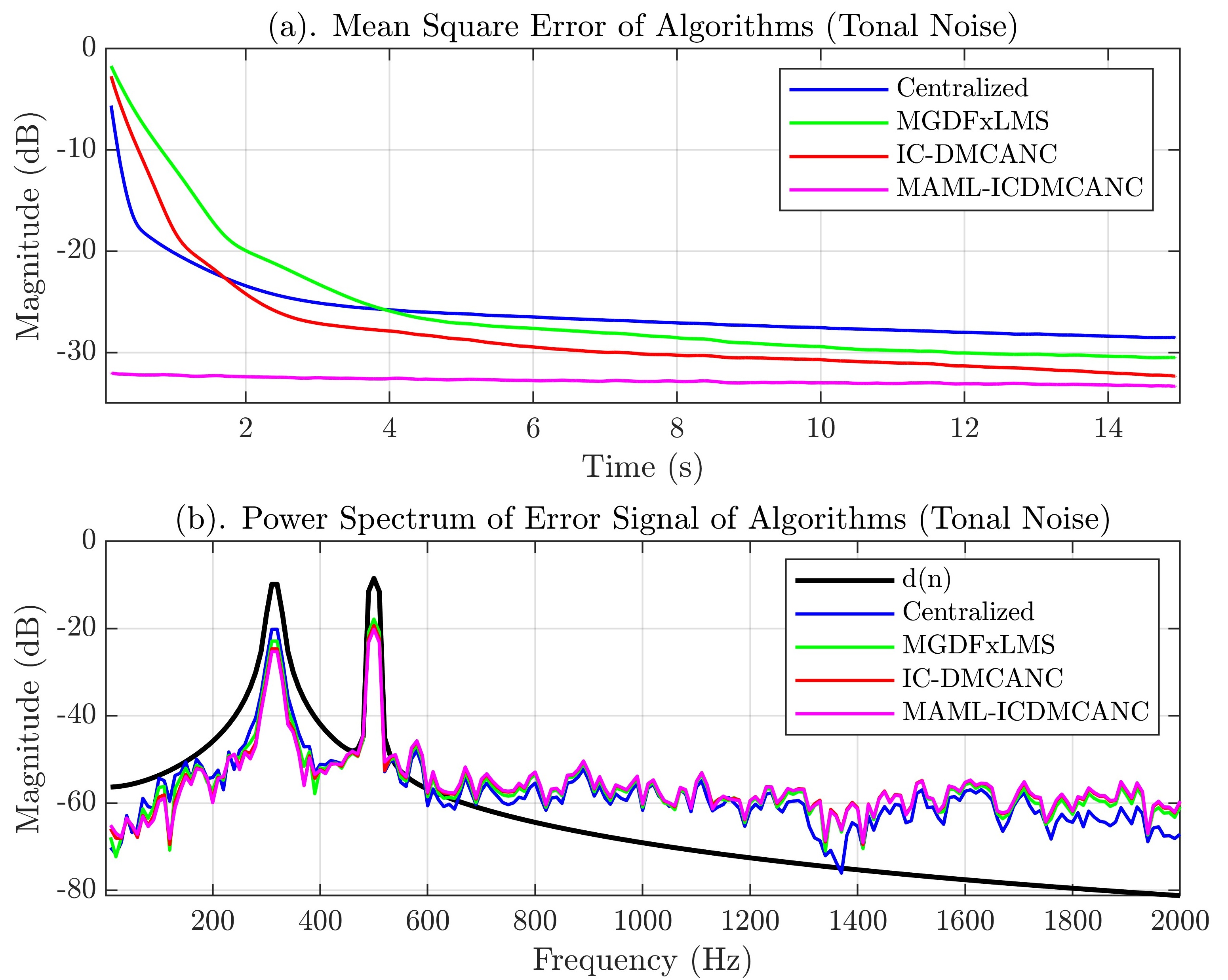}
    \caption{Convergence performance of different algorithms for tonal noise with ANC node 1: (a). Mean square error of different algorithms. (b). Power spectrum of error signals of different algorithms. }
    \label{fig_power1}
\end{figure}
A tonal noise consisting of $315$ Hz and $500$ Hz components was used for evaluation. The step size for all algorithms was set to $1\times 10^{-8}$. The mean square error (MSE) curves of the six nodes are shown in Fig.~\ref{fig_mse1}. It can be observed that the proposed method achieves the fastest convergence across all nodes. IC-DMCANC and centralized exhibit comparable convergence speed, both outperforming MGDFxLMS.The superior convergence performance of the proposed method is attributed to the MAML-based initialization, which captures prior acoustic characteristics before online adaptation. Due to differences in local acoustic environments, the steady-state performance varies slightly across nodes; however, consistent convergence acceleration is observed.

Fig.~\ref{fig_power1} presents the MSE and power spectrum of the error signals. The power spectrum was computed over the final $3$ seconds of the control process. As shown in Fig.~\ref{fig_power1}(b), all methods achieve comparable steady-state attenuation, while Fig.~\ref{fig_power1}(a) demonstrates the faster convergence of the proposed method.

\subsection{Broadband noise cancellation}
\begin{figure}[h!]
    \centering
    \includegraphics[width=0.8\linewidth,height = 9cm]{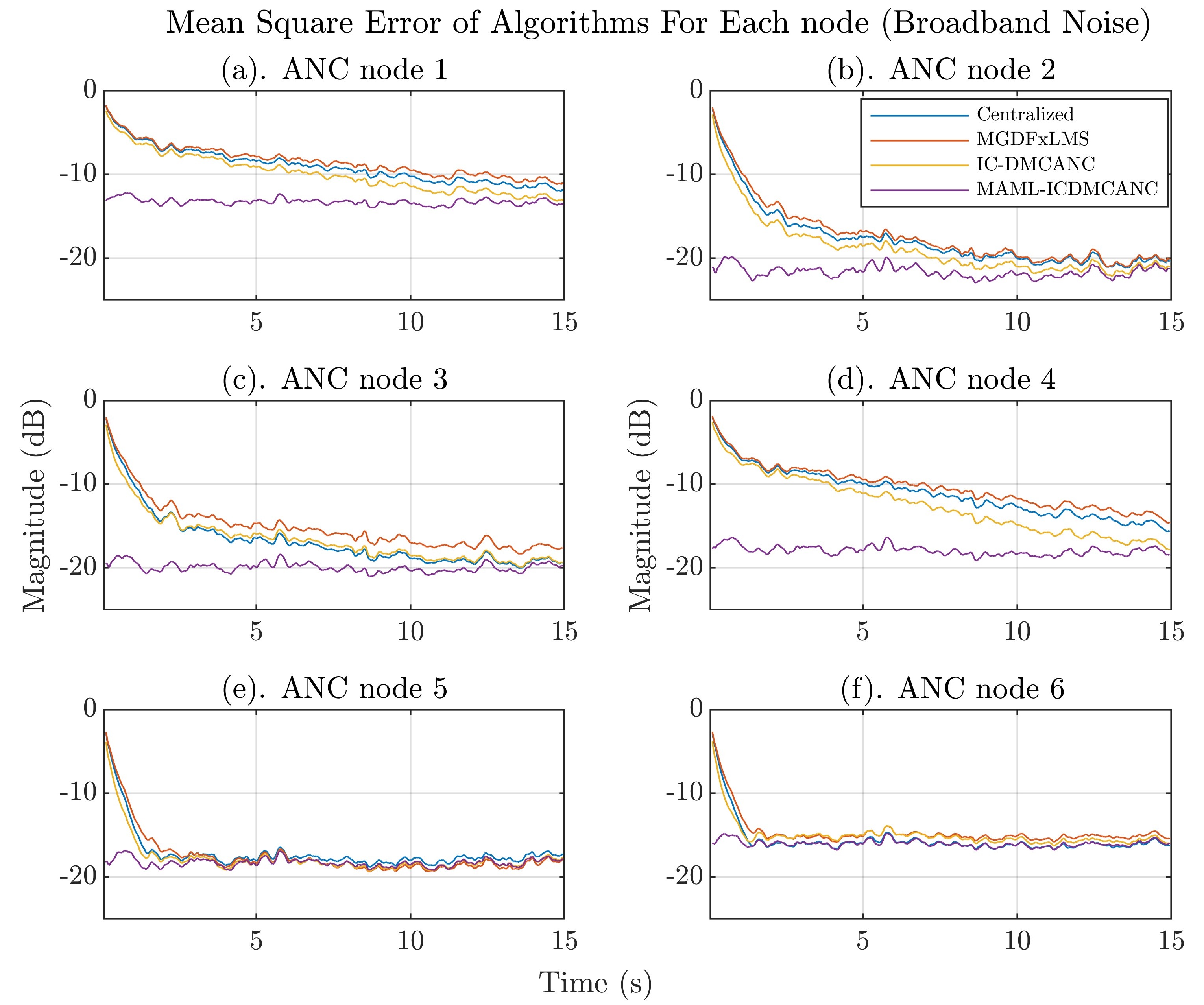}
    \caption{Mean square error of different algorithms for braobdand noise ($200-800$Hz) with each ANC node: (a). Mean square error in ANC node 1. (b).  Mean square error in ANC node 2. (c).  Mean square error in ANC node 3. (d).  Mean square error in ANC node 4. (e).  Mean square error in ANC node 5. (f). Mean square error in ANC node 6.}
    \label{fig_mse2}
\end{figure}

Broadband noise with the frequency range of $200$–$800$ Hz was used for evaluation. The step size was set to $3 \times 10^{-7}$ for all algorithms. The MSE curves for each node are shown in Fig.~\ref{fig_mse2}. Similar to the tonal noise case, the proposed method achieves the fastest convergence, while the other methods exhibit comparable convergence speed and similar steady-state performance.

\begin{figure}[h!]
    \centering
    \includegraphics[width=0.8\linewidth,height = 9cm]{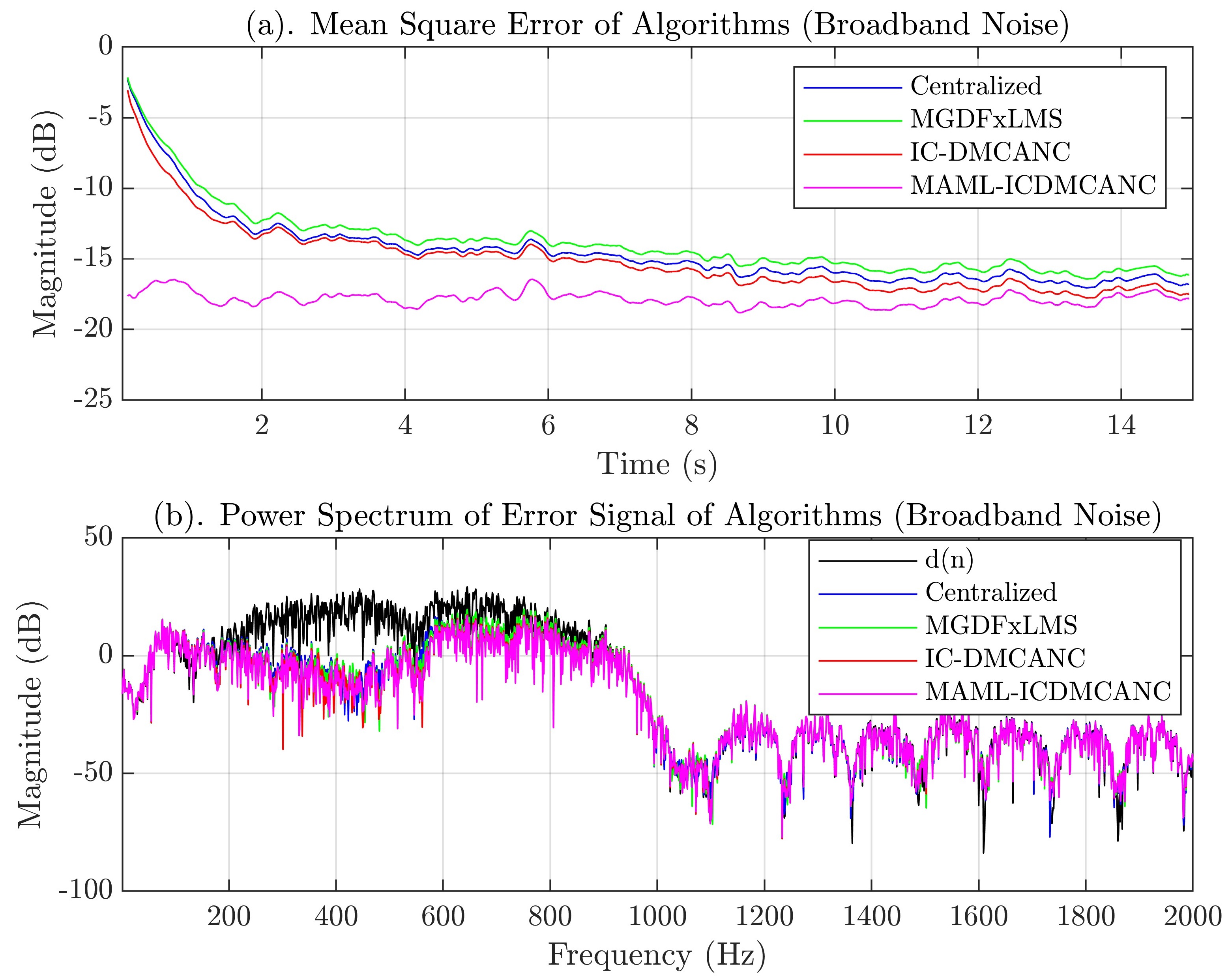}
    \caption{ Convergence performance of different algorithms for braobdand noise ($200-800$Hz) with ANC node 1: (a). Mean square error of different algorithms. (b). Power spectrum of error signals of different algorithms. }
    \label{fig_power2}
\end{figure}

Fig.~\ref{fig_power2} shows the convergence behavior and power spectrum for ANC node 1. All algorithms effectively attenuate noise within the target frequency band.

\subsection{Real noise cancellation}
In this simulation, a real recorded compressor noise was used. The step size settings were consistent with the previous simulations. The results indicate that the proposed method achieves the fastest convergence among all the compared methods.

\begin{figure}[h!]
    \centering
    \includegraphics[width=0.8\linewidth,height = 9cm]{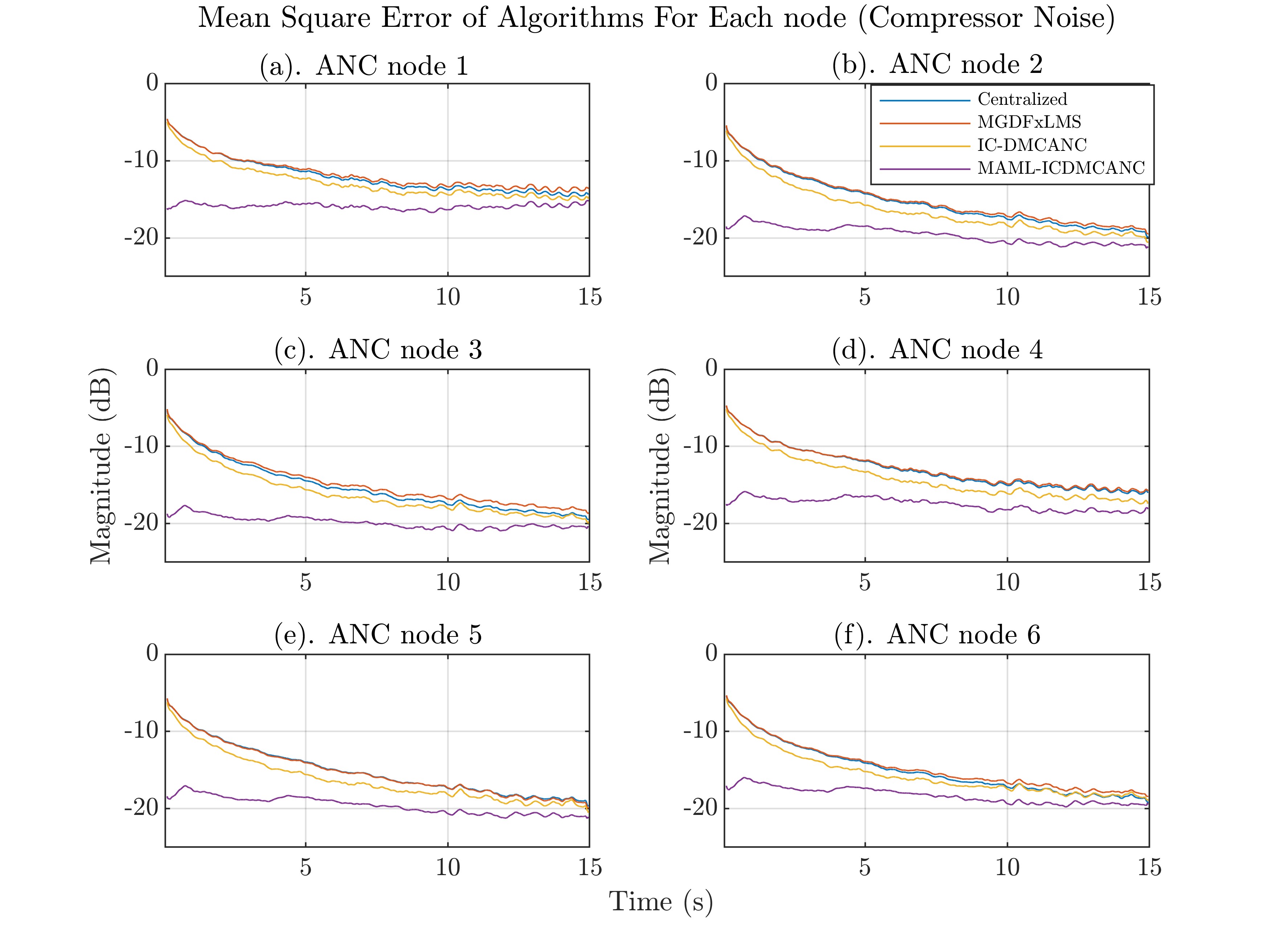}
    \caption{Mean square error of different algorithms for real recorded noise  with each ANC node: (a). Mean square error in ANC node 1. (b).  Mean square error in ANC node 2. (c).  Mean square error in ANC node 3. (d).  Mean square error in ANC node 4. (e).  Mean square error in ANC node 5. (f). Mean square error in ANC node 6.}
    \label{fig_mse3}
\end{figure}

\begin{figure}[h!]
    \centering
    \includegraphics[width=0.8\linewidth,height = 9cm]{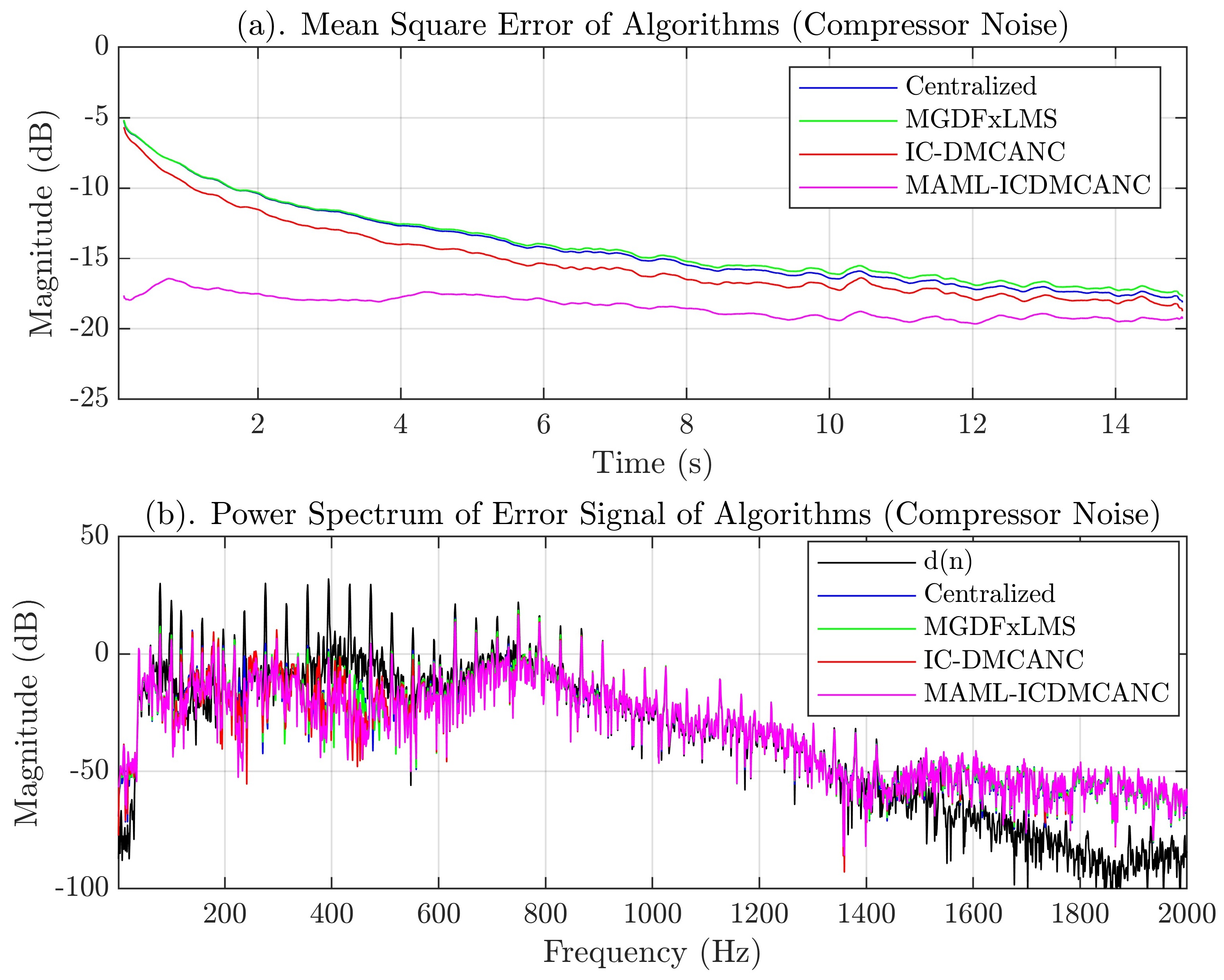}
    \caption{ Convergence performance of different algorithms real recorded noise with ANC node 1: (a). Mean square error of different algorithms. (b). Power spectrum of error signals of different algorithms. }
    \label{fig_power3}
\end{figure}

Fig.~\ref{fig_mse3} and Fig.~\ref{fig_power3} illustrate the MSE and power spectrum of the error signals, respectively. As shown in Fig.~\ref{fig_mse3}, the proposed method suppresses the noise from the early stage of the control process, whereas other methods begin to achieve noticeable attenuation after approximately $10$ seconds. Fig.~\ref{fig_power3} further confirms that the proposed method effectively attenuates noise within the target frequency range.

\section{CONCLUSION}\label{Sec_4}
This paper proposed a model-agnostic meta-learning (MAML)-based initialization framework for distributed multichannel active noise control with intermittent communication (IC-DMCANC). By leveraging heterogeneous acoustic information across different nodes and environments, the proposed method learned a generalized initialization that significantly improved the convergence behavior of adaptive control filters. Unlike conventional DMCANC approaches that relied on zero or random initializations, the proposed framework enabled each node to start from a more informative initial point, thereby accelerating adaptation and enhancing inter-node collaboration. The method was fully compatible with existing distributed ANC architectures and could be seamlessly integrated into practical systems without modifying the underlying control structure. Simulations under different noise conditions demonstrated that the proposed method achieved faster convergence performance compared with conventional methods. These findings highlighted the importance of initialization in DMCANC and validated the effectiveness of MAML in addressing this previously overlooked problem.
\section*{Acknowledgements}
This research is supported by the Chinese Academy of Sciences Talent Introduction Program for Young Researchers (Grant Number: 552025000180).

\bibliographystyle{unsrt}
\bibliography{a} 

@article{lam2018active,
  title={Active control of sound through full-sized open windows},
  author={Lam, Bhan and Shi, Chuang and Shi, Dongyuan and Gan, Woon-Seng},
  journal={Building and Environment},
  volume={141},
  pages={16--27},
  year={2018},
  publisher={Elsevier}
}

@article{Kuo1999ANC,
   author = {Kuo, Sen M and Morgan, Dennis R},
   title = {Active noise control: a tutorial review},
   journal = {Proceedings of the IEEE},
   volume = {87},
   number = {6},
   pages = {943-973},
   ISSN = {0018-9219},
   year = {1999},
   type = {Journal Article}
}

@article{Ji2025SBWCFXLMS,
   author = {Ji, Junwei and Shi, Dongyuan and Luo, Zhengding and Wang, Boxiang and Gan, Woon-Seng},
   title = {Self-Boosted Weight-Constrained FxLMS: A Robustness Distributed Active Noise Control Algorithm Without Internode Communication},
   journal = {IEEE Signal Processing Letters},
   pages = {1-5},
   ISSN = {1070-9908},
   DOI = {10.1109/lsp.2025.3588421},
   url = {https://dx.doi.org/10.1109/lsp.2025.3588421},
   year = {2025},
   type = {Journal Article}
}

@article{shi2022selective,
  title={Selective fixed-filter active noise control based on convolutional neural network},
  author={Shi, Dongyuan and Lam, Bhan and Ooi, Kenneth and Shen, Xiaoyi and Gan, Woon-Seng},
  journal={Signal Processing},
  volume={190},
  pages={108317},
  year={2022},
  publisher={Elsevier}
}

@inproceedings{Ji2023Distributed,
   author = {Ji, Junwei and Shi, Dongyuan and Luo, Zhengding and Shen, Xiaoyi and Gan, Woon-Seng},
   title = {A practical distributed active noise control algorithm overcoming communication restrictions},
   booktitle = {2023 Proc. - ICASSP IEEE Int. Conf. Acoust. Speech Signal Process.},
   publisher = {IEEE},
   year = {2023},
   pages = {1-5},
   ISBN = {1728163277},
   type = {Conference Proceedings}
}

@article{Ji2025MGDFXLMS,
   author = {Ji, Junwei and Shi, Dongyuan and Gan, Woon-Seng},
   title = {Mixed-Gradients Distributed Filtered Reference Least Mean Square Algorithm – A Robust Distributed Multichannel Active Noise Control Algorithm},
   journal = {IEEE Transactions on Audio, Speech and Language Processing},
   volume = {33},
   pages = {1563-1575},
   ISSN = {2998-4173},
   DOI = {10.1109/taslpro.2025.3552932},
   url = {https://dx.doi.org/10.1109/taslpro.2025.3552932},
   year = {2025},
   type = {Journal Article}
}

@article{Li2023Distributed,
   author = {Li, Tianyou and Lian, Siyuan and Zhao, Sipei and Lu, Jing and Burnett, Ian S.},
   title = {Distributed Active Noise Control Based on an Augmented Diffusion FxLMS Algorithm},
   journal = {IEEE/ACM Transactions on Audio, Speech, and Language Processing},
   pages = {1-15},
   ISSN = {2329-9290},
   DOI = {10.1109/taslp.2023.3261742},
   url = {https://dx.doi.org/10.1109/taslp.2023.3261742},
   year = {2023},
   type = {Journal Article}
}

@inproceedings{shen2024survey,
  title={A survey of integrating wireless technology into active noise control},
  author={Shen, Xiaoyi and Shi, Dongyuan and Luo, Zhengding and Ji, Junwei and Gan, Woon-Seng},
  booktitle={INTER-NOISE and NOISE-CON Congress and Conference Proceedings},
  volume={270},
  pages={1624--1637},
  year={2024},
  organization={Institute of Noise Control Engineering}
}

@inproceedings{shi2018partial,
  title={A partial-update minimax algorithm for practical implementation of multi-channel feedforward active noise control},
  author={Shi, Chaung and Kajikawa, Yoshinobu},
  booktitle={2018 16th International Workshop on Acoustic Signal Enhancement (IWAENC)},
  pages={1--15},
  year={2018},
  organization={IEEE}
}

@article{Li2023AugmentedDiffusion,
   author = {Li, Tianyou and Rao, Li and Zhao, Sipei and Duan, Hongji and Lu, Jing and Burnett, Ian S.},
   title = {An augmented diffusion algorithm with bidirectional communication for a distributed active noise control system},
   journal = {The Journal of the Acoustical Society of America},
   volume = {154},
   number = {6},
   pages = {3568-3579},
   ISSN = {0001-4966},
   DOI = {10.1121/10.0022573},
   url = {https://dx.doi.org/10.1121/10.0022573},
   year = {2023},
   type = {Journal Article}
}

@article{Chu2020DiffusionANC,
   author = {Chu, Y. J. and Mak, C. M. and Zhao, Y. and Chan, S. C. and Wu, M.},
   title = {Performance analysis of a diffusion control method for ANC systems and the network design},
   journal = {Journal of Sound and Vibration},
   volume = {475},
   pages = {115273},
   ISSN = {0022-460X},
   DOI = {https://doi.org/10.1016/j.jsv.2020.115273},
   url = {https://www.sciencedirect.com/science/article/pii/S0022460X20301048},
   year = {2020},
   type = {Journal Article}
}

@article{zhou2025distributed,
  title={Distributed Active Noise Control Robust to Impulsive Interference},
  author={Zhou, Yang and Zhao, Haiquan and Liu, Dongxu and Guo, Xinnian},
  journal={IEEE Sensors Journal},
  year={2025},
  publisher={IEEE}
}

@book{Elliott2001SPAC,
   author = {Elliott, S. J.},
   title = {Signal processing for active control},
   publisher = {Academic},
   address = {San Diego, Calif.},
   note = {Includes bibliographical references and index.},
   ISBN = {9780080517131},
   year = {2001},
   type = {Book}
}

@inproceedings{finn2017model,
  title={Model-agnostic meta-learning for fast adaptation of deep networks},
  author={Finn, Chelsea and Abbeel, Pieter and Levine, Sergey},
  booktitle={International conference on machine learning},
  pages={1126--1135},
  year={2017},
  organization={PMLR}
}

@inproceedings{zintgraf2019fast,
  title={Fast context adaptation via meta-learning},
  author={Zintgraf, Luisa and Shiarli, Kyriacos and Kurin, Vitaly and Hofmann, Katja and Whiteson, Shimon},
  booktitle={International Conference on Machine Learning},
  pages={7693--7702},
  year={2019},
  organization={PMLR}
}

@article{shi2024behind,
  title={What is behind the meta-learning initialization of adaptive filter?—A naive method for accelerating convergence of adaptive multichannel active noise control},
  author={Shi, Dongyuan and Gan, Woon-seng and Shen, Xiaoyi and Luo, Zhengding and Ji, Junwei},
  journal={Neural Networks},
  pages={106145},
  year={2024},
  publisher={Elsevier}
}

@article{kajikawa2012recent,
  title={Recent advances on active noise control: open issues and innovative applications},
  author={Kajikawa, Yoshinobu and Gan, Woon-Seng and Kuo, Sen M},
  journal={APSIPA Transactions on Signal and Information Processing},
  volume={1},
  year={2012},
  publisher={Cambridge University Press}
}

@book{hansen1999understanding,
  title={Understanding active noise cancellation},
  author={Hansen, Colin N},
  year={1999},
  publisher={CRC Press}
}

@article{chang2020active,
  title={Active noise control for centrifugal and axial fans},
  author={Chang, Cheng-Yuan and Liu, Xiu-Wei and Kuo, Sen M and others},
  journal={Noise Control Engineering Journal},
  volume={68},
  number={6},
  pages={490--500},
  year={2020},
  publisher={Institute of Noise Control Engineering}
}

@article{SHEN2022117300,
title = {Multi-channel wireless hybrid active noise control with fixed-adaptive control selection},
journal = {Journal of Sound and Vibration},
volume = {541},
pages = {117300},
year = {2022},
issn = {0022-460X},
doi = {https://doi.org/10.1016/j.jsv.2022.117300},
url = {https://www.sciencedirect.com/science/article/pii/S0022460X22004837},
author = {Xiaoyi Shen and Woon-Seng Gan and Dongyuan Shi}
}

@article{shen2022adaptive,
  title={Adaptive-gain algorithm on the fixed filters applied for active noise control headphone},
  author={Shen, Xiaoyi and Shi, Dongyuan and Gan, Woon-Seng and Peksi, Santi},
  journal={Mechanical Systems and Signal Processing},
  volume={169},
  pages={108641},
  year={2022}
}

@book{haykin2002adaptive,
  title={Adaptive filter theory},
  author={Haykin, Simon S},
  year={2002},
  publisher={Pearson Education India}
}

@article{gan2023practical,
  title={Practical Active Noise Control: Restriction of Maximum Output Power},
  author={Gan, Woon-Seng and Shi, Dongyuan and Shen, Xiaoyi},
  journal={arXiv preprint arXiv:2307.10913},
  year={2023}
}

@inproceedings{shen2023momentum,
  title={A Momentum Two-Gradient Direction Algorithm with Variable Step Size Applied to Solve Practical Output Constraint Issue for Active Noise Control},
  author={Shen, Xiaoyi and Shi, Dongyuan and Luo, Zhengding and Ji, Junwei and Gan, Woon-Seng},
  booktitle={ICASSP 2023-2023 IEEE International Conference on Acoustics, Speech and Signal Processing (ICASSP)},
  pages={1--5},
  year={2023},
  organization={IEEE}
}

@inproceedings{luo2023deep,
  title={Deep Generative Fixed-Filter Active Noise Control},
  author={Luo, Zhengding and Shi, Dongyuan and Shen, Xiaoyi and Ji, Junwei and Gan, Woon-Seng},
  booktitle={ICASSP 2023-2023 IEEE International Conference on Acoustics, Speech and Signal Processing (ICASSP)},
  pages={1--5},
  year={2023},
  organization={IEEE}
}

@article{shi2021fast,
  title={Fast adaptive active noise control based on modified model-agnostic meta-learning algorithm},
  author={Shi, Dongyuan and Gan, Woon-Seng and Lam, Bhan and Ooi, Kenneth},
  journal={IEEE Signal Processing Letters},
  volume={28},
  pages={593--597},
  year={2021},
  publisher={IEEE}
}

@article{shi2023transferable,
  title={Transferable latent of cnn-based selective fixed-filter active noise control},
  author={Shi, Dongyuan and Gan, Woon-Seng and Lam, Bhan and Luo, Zhengding and Shen, Xiaoyi},
  journal={IEEE/ACM Transactions on Audio, Speech, and Language Processing},
  volume={31},
  pages={2910--2921},
  year={2023},
  publisher={IEEE}
}

@article{SHEN2025112415,
title = {Multi-channel adjoint least mean square algorithm with momentum factor applied on active noise control},
journal = {Mechanical Systems and Signal Processing},
volume = {228},
pages = {112415},
year = {2025},
issn = {0888-3270},
doi = {https://doi.org/10.1016/j.ymssp.2025.112415},
url = {https://www.sciencedirect.com/science/article/pii/S0888327025001165},
author = {Xiaoyi Shen and Yu Guo and Junwei Ji and Dongyuan Shi and Woon-Seng Gan}
}

@article{JI2026114024,
title = {Implementation of distributed multichannel active noise control with intermittent communication and coprocessor assisted data combination},
journal = {Mechanical Systems and Signal Processing},
volume = {248},
pages = {114024},
year = {2026},
issn = {0888-3270},
doi = {https://doi.org/10.1016/j.ymssp.2026.114024},
url = {https://www.sciencedirect.com/science/article/pii/S0888327026001810},
author = {Junwei Ji and Dongyuan Shi and Xiaoyi Shen and Woon-Seng Gan and Jie Chen and Jun Yang}
}

@ARTICLE{11021385,
  author={Shen, Xiaoyi and Shi, Dongyuan and Gan, Woon-Seng},
  journal={IEEE Signal Processing Letters}, 
  title={Data-Driven Method to Accelerate Convergence of Adaptive Hybrid Active Noise Control: Two-Stage Model-Agnostic Meta-Learning}, 
  year={2025},
  volume={32},
  number={},
  pages={2654-2658},
  doi={10.1109/LSP.2025.3576033}}

\end{document}